\documentclass{article}
\usepackage{spconf,amsmath,graphicx}
\usepackage[flushleft]{threeparttable}
\usepackage{multirow}
\usepackage{makecell}
\usepackage{pifont}
\usepackage{amsmath}
\usepackage{booktabs} 
\usepackage{algorithm}
\usepackage{algpseudocode}
\usepackage{graphicx}
\usepackage{enumitem}
\usepackage{subcaption}
\usepackage{afterpage}

\title{Dynamic ASR pathways: An Adaptive Masking Approach towards efficient pruning of a multilingual ASR model}
\name{\makecell{Jiamin Xie$^{\ast\thanks{{$^{\ast}$} Work done while Jiamin Xie was an intern at Meta AI.},1}$ \footnote{a}, Ke Li$^2$, Jinxi Guo$^2$, Andros Tjandra$^2$, Yuan Shangguan$^2$, \\ Leda Sari$^2$, Chunyang Wu$^2$, Junteng Jia$^2$, Jay Mahadeokar$^2$, Ozlem Kalinli$^2$}}
\address{$^1$Center for Robust Speech Systems (CRSS), University of Texas at Dallas, USA\\ $^2$Meta AI, USA}
\email{jiamin.xie@utdallas.edu, kli26@meta.com}

\begin{document}
\ninept
\maketitle
\begin{abstract}
Neural network pruning offers an effective method for compressing a multilingual automatic speech recognition (ASR) model with minimal performance loss. However, it entails several rounds of pruning and re-training needed to be run for each language. In this work, we propose the use of an adaptive masking approach in two scenarios for pruning a multilingual ASR model efficiently, each resulting in sparse monolingual models or a sparse multilingual model (named as \textit{Dynamic ASR Pathways}). Our approach dynamically adapts the sub-network, avoiding premature decisions about a fixed sub-network structure. We show that our approach outperforms existing pruning methods when targeting sparse monolingual models. Further, we illustrate that \textit{Dynamic ASR Pathways} jointly discovers and trains better sub-networks (pathways) of a single multilingual model by adapting from different sub-network initializations, thereby reducing the need for language-specific pruning.
\end{abstract}
\section{Introduction}
Automatic speech recognition (ASR) has become a key feature in smart devices, serving a diverse customer base \cite{shangguan2021dissecting, he2019streaming, gao2021extremely}. For a successful on-device deployment, the ASR model must operate within the storage and computational constraints while delivering an optimal performance. Furthermore, the ASR model needs to support multiple languages \cite{massive_mutlilingual, tjandra2023massively} to interact with users worldwide. Neural network pruning \cite{imp,progressive,frankle2018lottery} is an effective technique for reducing the size of an ASR model with minimal performance loss \cite{shangguan2019optimizing, narang2017exploring}. However, the pruning process, such as Iterative Magnitude Pruning (IMP) \cite{imp, renda2020comparing} and Lottery Ticket Hypothesis (LTH) \cite{frankle2018lottery}, involves multiple iterations of pruning and re-training to achieve the best performance. The pruning step identifies a task-specific sub-network within the original dense neural network. Subsequently, the re-training step trains this sub-network with task-specific data, mitigating the performance loss introduced in pruning. This iterative process continues until the target sparsity level is reached.

Pruning a multilingual ASR model presents specific challenges. When pruning a pre-trained dense multilingual ASR model, it can result in two scenarios, as discussed in \cite{yang2023learning}. In the first scenario, the model is fine-tuned and pruned for each language separately, resulting in multiple language-specific sparse models. While this approach optimizes performance in each language, it can increase storage requirements due to maintaining different monolingual models. In the second scenario, the multilingual model is fine-tuned and pruned using a multilingual dataset, creating a single sparse model by finding a language-agnostic pruning mask. While multilingual training can promote knowledge transfer across languages \cite{datta2020language, ogueji-etal-2022-intriguing}, data imbalance \cite{kannan19_interspeech, winata21_interspeech} may cause performance degradation in some languages when training a single language-agnostic sub-network. Mixing languages in a training batch can also create conflicts in weight updates with different languages fighting for model capacity, known as the negative interference effect \cite{yu2020gradient, shaham-etal-2023-causes}, making it challenging to identify an optimal language-agnostic sub-network. A recent study \cite{yang2023learning} proposes to train language-specific sub-networks (referred to as pathways) jointly within the original dense multilingual model instead of training a language-agnostic sub-network. This method employs monolingual data in a batch to fine-tune the respective pathway without interference from other languages. As these pathways overlap, the weights are updated either in a language-specific or a language-agnostic manner, surpassing the performance of language-agnostic methods. However, a drawback of the pathways method is acquiring each pathway in a separate stage that performs monolingual training and pruning, incurring a computational cost that scales linearly with the number of languages. These pathways, once obtained, remain fixed throughout the training process, lacking adaptation to the multilingual data.


In this study, we introduce an adaptive masking approach for adapting language-specific sub-networks in monolingual or multilingual pruning situations. Our proposed method re-evaluates the pruning mask dynamically during training, allowing the sub-network to align better with the training data comparing to a fixed masking approach. We first assess the benefit of applying this technique to the monolingual case, obtaining sparse monolingual ASR models. We then prune and adapt pathways by employing our approach in multilingual training, evaluating the performance of a jointly fine-tuned and pruned multilingual ASR model.
\begin{figure}[t]
  \centering
  \includegraphics[width=24em, height=19em]{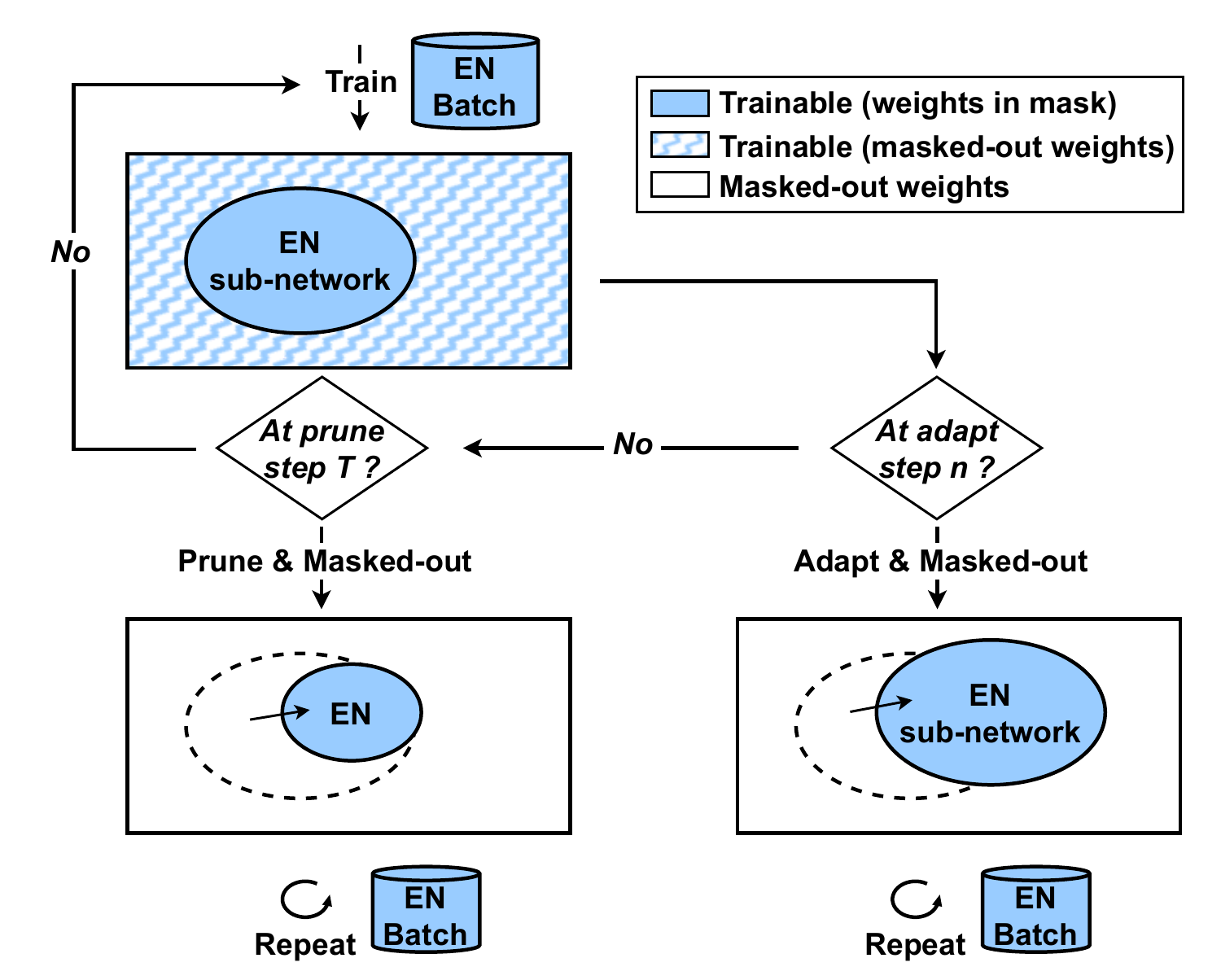}
  \caption{Flowchart of the training and pruning process with adaptive masking enabled for monolingual data}
  \label{variable_monolingual}
\end{figure}
\section{Related Works}
\textbf{Multilingual Training.} The concept of training sub-networks was proposed in the context of multi-task learning \cite{Sun_Shao_Li_Liu_Yan_Qiu_Huang_2020} and has since found applications in multilingual training \cite{language-adaptive,lin-etal-2021-learning,foroutan-etal-2022-discovering}. This approach has demonstrated an efficacy across various domains, including self-supervised learning (SSL) \cite{language-adaptive}, machine translation \cite{lin-etal-2021-learning}, and language modeling \cite{foroutan-etal-2022-discovering}. Our research builds upon a recent study \cite{yang2023learning} that emphasized the effectiveness of training language-specific sub-networks for the supervised multilingual ASR task. 

\noindent\textbf{Adaptive Pruning.}
Previous research on adaptive pruning can be broadly categorized based on whether the pruning masks are made trainable. One approach \cite{fu2022losses} involves fine-tuning the trainable pruning masks on downstream tasks while keeping the original model weights fixed, demonstrating performance improvements over traditional fine-tuning methods. Another approach \cite{lai2021parp} focuses on re-learning the pruned weights by lifting the mask during training, allowing adjustments to the pruning mask without learning it directly. This technique was applied to fine-tune a multilingual speech SSL model for monolingual ASR tasks. In contrast to the latter approach, our study applies adaptive masking to the supervised multilingual ASR task with structured pruning, introducing novel strategies for attaining and adapting sub-networks during multilingual training.
\vspace{-0.5em}
\section{Methodology}
We first recap the concept of pruning (Section 3.1). We then illustrate current pruning methods that are foundational to our proposed approach (Section 3.2). Finally, we described our adaptive masking approach for monolingual and multilingual pruning (Section 3.3).

\subsection{Pruning recap}
For a dense neural network $f(x;\theta_{0})$ trained with 
input sample $x$ and parameters $\theta_{0}$, we denote a sub-network $f(x; m \odot \theta_{0})$ with a binary pruning mask $m$ and the element-wise product $\odot$. The pruning goal is to identify the sub-network $f(x; m\odot\theta)$ through additional training, where $\theta$ can be the parameters obtained at any stage of training. We consider a progressive pruning schedule \cite{progressive}, where pruning starts from a low sparsity and incrementally steps up to the target sparsity.

\subsection{Current pruning methods}
\subsubsection{IMP, LTH, and LAP}
The iterative magnitude pruning (IMP) method \cite{imp} involves fine-tuning a dense model for a specified number of steps denoted as $T$ while making pruning decisions based on the magnitude of weights. Here, the magnitude of a weight reflects its significance to the task, with larger values indicating higher importance. For structured pruning, we use the block-wise scheme similar to that in \cite{yang2022omni}, following a block pattern of 8 $\times$ 1. This pattern implies that eight consecutive weights within a column are pruned simultaneously. We evaluate magnitudes by the L2 norm of a weight block. To initiate the IMP procedure, we initialize model parameters $\theta$ with pre-trained dense weights $\theta_{0}$ and set the binary pruning mask $m$ to all ones $\mathbf{1}$, where $m \in \{0,1\}^{|\theta|}$.\begin{figure}[t]
  \centering
  \includegraphics[width=24em, height=19em]{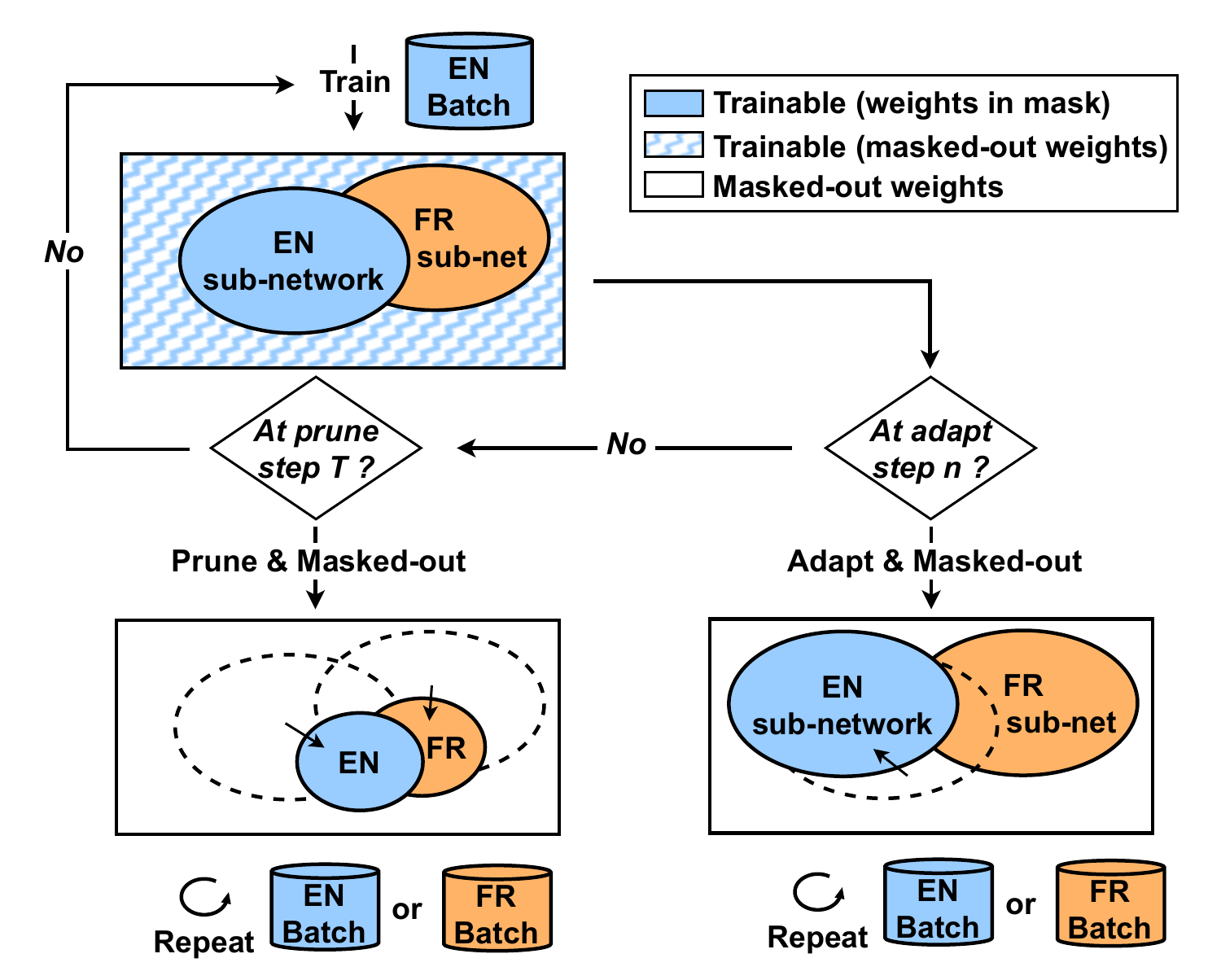}
  \caption{Flowchart of the training and pruning process with adaptive masking enabled for multilingual data}
  \label{variable_multilingual}
\end{figure} The IMP procedure is illustrated as follows:

\noindent\textbf{Repeat}
\begin{enumerate}
    \item Train $f(x; m\odot\theta)$ for $T$ steps, resulting in  $f(x; m \odot \theta_{T})$.
    \item Prune $p\%$ of total weights from $m \odot \theta_{T}$ that has the smallest magnitudes. Setting the pruned positions in $m$ to 0.
    \item Assign $\theta_{T}$ to $\theta$ for the next iteration
\end{enumerate}
\noindent\textbf{Until} $m$ reaches the target sparsity

\noindent We note the property of a pruning mask depends on the training data. When monolingual data is used in IMP, this procedure yields a language-specific pruning mask $m_{l}$ for a language $l$. For multilingual data, it results in a language-agnostic pruning mask, referred to as language-agnostic pruning (LAP). Importantly, the pruning mask remains fixed at any step within the $T$ training steps, suggesting the pruning decision is irreversible.
\begin{table*}[h!]
    \centering
    \caption{WER (\%) results on the MLS test set, pruning a dense multilingual ASR model. The proposed approach allows the mask to change in training and is compared to other pruning methods for monolingual training scenario.}
    \resizebox{1.95\columnwidth}{!}{
    \begin{tabular}{c | c | c | c | c | c  c c c | c }
        \toprule
        Stage & Model & \makecell{Mask can\\ change?} & Sparsity & \makecell{Monolingual or \\ Multilingual training?} & EN & FR & IT & NL & Avg. \\
        \midrule
        \midrule
        Ref. & 56M Dense & / & 0\% & Monolingual & 12.15 & 16.00 & 27.62 & 23.23 & 19.75 \\
        \midrule
        (1) & 187M Dense & / & 0\% & Multilingual & 12.91 & 10.90 & 16.94 & 17.56 & 14.58 \\
        \midrule
        \multirow{3}{*}{\centering (2)} & LAP & No & 70\% & Multilingual & 13.82 & 11.98 & 27.71 & 19.32 & 18.21 \\
        & IMP & No & 70\% & Monolingual & 10.74 & 11.26 & 17.90 & 18.38 & 14.57 \\
        & LTH & No & 70\% & Monolingual & 10.80 & 10.38 & 18.44 & 17.48 & 14.28 \\
        \midrule
         \multirow{2}{*}{\centering (3)} & ASR Pathways (IMP-70\%) & No & 70\% & Multilingual & 11.15 & 10.68 & 17.53 & 16.90 & 14.06 \\
         & ASR Pathways (LTH-70\%) & No & 70\% & Multilingual & 11.39 & 10.20 & 17.58 & 15.84 & 13.75 \\
         \midrule
         \multirow{2}{*}{\makecell{(2) \\ \textit{Proposed}}} & IMP & Yes & 70\% & Monolingual & \textbf{10.07} & 10.90 & 17.21 & 16.98 & 13.79 \\
         & LTH & Yes & 70\% & Monolingual & 10.54 & \textbf{9.91} & \textbf{17.06} & \textbf{16.63} & \textbf{13.53} \\
         \bottomrule
    \end{tabular}
     }
    \label{baselines}
\end{table*}

The lottery ticket hypothesis (LTH) method \cite{frankle2018lottery} modifies the Step 3 of the IMP procedure by assigning the pre-trained dense weights $\theta_{0}$ to $\theta$ instead of $\theta_{T}$, referred to as a re-winding step. It assumes that a sub-network capable of achieving performance similar to the original dense network exists within the original dense architecture. Therefore, the LTH method leads to the identification of a sub-network embedded within the original dense model weights.
\subsubsection{ASR Pathways}
The ASR Pathways \cite{yang2023learning} provides a method to fine-tune a multilingual ASR model using the language-specific sub-networks (or pathways) identified through IMP, LTH, or other pruning methods. These sub-networks are attained at the target sparsity level and remain fixed throughout training. The mini-batch is configured as monolingual, while the training includes a mixture of languages across mini-batches. This setup ensures that each mini-batch activates one pathway and updates the weights underlying the pruning mask of this pathway, denoted as $m_{l} \odot \theta$. Since each language-specific sub-network is a part of the original dense multilingual model and gets fine-tuned together, the training process results in a final sparse multilingual ASR model.

\subsection{The adaptive masking approach}

\subsubsection{Monolingual pruning}
We propose an adaptive masking approach for monolingual pruning, yielding a language-specific pruning mask adapted with the data. We illustrate this approach as a flowchart shown in Figure \ref{variable_monolingual}. Within the framework of the IMP procedure, we introduce a mask adaptation step denoted as $n$ (where $n < T$). During the adaptation step, we re-evaluate the sub-network configuration (\textit{adapt}) by pruning from all weights in $\theta_{n}$ with a portion $p$\% that maintains the sparsity level of the current pruning mask. Next, we prune ``softly" by setting the pruned weights to zero, denoted as $(\mathbf{1}-m)\ \odot\  \theta_n$, and make them trainable (\textit{masked-out}). Since the masked-out weights receive updates from training, they can form new connections within the network and reveal an optimal configuration of the sub-network as the training evolves. For the pruning step, we simply raise the sparsity level and prune from all weights in $\theta_{n}$ as opposed to pruning from weights in $m \odot \theta_{T}$ in the IMP procedure.

\subsubsection{Multilingual pruning}
We propose an adaptive masking approach for multilingual pruning based on the pathways training method described in \cite{yang2023learning}, named \textit{Dynamic ASR Pathways}. We use a similar adaptation step to the monolingual pruning and illustrate it in a flowchart shown in Figure \ref{variable_multilingual}. When a mini-batch in language $z$ is processed, we train the sub-network of this language $z$ and a ``residual" sub-network, excluding other language-specific sub-networks. Given a language set $L$ representing all languages in the data, we denote this pruning mask as,
\begin{equation}
    m_{z,r} = m_{z} \cup (\mathbf{1} - \cup_{l\ in\ L, l \neq z} m_{z})
\end{equation}
During the adaption step, we re-evaluate the language-specific sub-network by pruning from weights in $m_{z,r} \odot \theta_{n}$ with its current sparsity level held. Since the adaptation step is monolingual, the newly adapted sub-network can become more language-specific compared to before. During the pruning step, we simultaneously prune sub-networks by pruning from weights in $m_{z,r} \odot \theta_{T}$, iterating over each language $z$ in the language set $L$. Because different languages would share the ``residual" sub-network depending on the data distribution, this pruning step promotes parameter sharing among sub-networks, compensating potential reductions in the adaptation step.


\section{Experimental Setup}
\label{sec:typestyle}
\subsection{Dataset}
We conduct our experiments using the multilingual Librispeech (MLS) \cite{mls} dataset, which consists of multilingual speech derived from audiobooks. Our study focuses on four languages: English (EN), French (FR), Italian (IT), and Dutch (NL), with respective training audio length of 44.7k hrs, 1.1k hrs, 0.2k hrs, 1.6k hrs.

\subsection{Implementation details}
We employ a streaming RNN-T model for the dense multilingual model, using 30 Emformer layers \cite{shi2021emformer} with 512 input dimensions, 2048 feed-forward dimensions, and encoder layers with convolutional blocks \cite{gulati2020conformer}. This model has about 180 million parameters. We utilize word pieces to recognize spoken words in all four languages, totaling 1548 items. For consistency, we use the same output layer size for all training setups. The learning rate schedule is tri-stage \cite{park2019specaugment} with a peak learning rate of 1e-3. For monolingual models, we conduct training for 100K, 80K, 50K, and 80K steps for EN, FR, IT, and NL, respectively. The multilingual pathway model undergoes training for 200K and 100K steps for IMP and LTH methods, respectively. We also conduct a bilingual experiment for the multilingual pathway models, where the training step is 80K. We employ an uniform data sampling scheme for the multilingual training when \textit{Dynamic ASR Pathways} method is compared, otherwise an non-uniform sampling scheme.  The prune step $T$ was set to be 8\% of the training step for each setup, with an adaption step $n$ of $100$ and a prune portion $p$ of 20\% across all experiments. Pruning was applied exclusively to linear layers in the encoder Eformer and the predictor LSTM layers, with a uniform sparsity across all prunable layers \cite{yang2022omni}. We apply group lasso regularization following \cite{liu2023learning}. We use 16 GPUs for monolingual training and 32 GPUs for multilingual training, with a per-GPU batch size of 28.  
\section{Results}
We first show baseline results from using current pruning methods (Section 5.1). We then compare the adaptive masking approach for monolingual pruning to its relative baseline (Section 5.2). Finally, we compare the adaptive masking approach for multilingual pruning (\textit{Dynamic ASR Pathways}) to the ASR Pathways baseline (Section 5.3). 

\begin{table}[H]
  \caption{WER (\%) results on the MLS test set, utilizing language-specific pruning masks. The proposed approach is compared to an existing method for the bilingual training case.}
  \label{t2}
  \centering
  \scalebox{0.85}{
  \begin{tabular}{ c | c | c | c | c c | c }
    \toprule
     Model & \makecell{Initialization} & \makecell{Mask\\ change?} & Sparsity & FR & NL & Avg. \\
    \midrule
    \midrule
    \makecell{ASR \\ Pathways} & LTH-70\% & No & 70\% & 10.73 & 16.23 & 13.48  \\
    \midrule
    \makecell{ASR \\ Pathways} & LAP-70\% & No & 70\% & 11.98 & 19.32 & 15.65 \\
    \midrule
    \multirow{3}{*}{\textit{\makecell{Dynamic \\ ASR \\ Pathways}}} & LTH-70\% & Yes & 70\% & 11.31 & 15.55 & 13.43 \\
    & LTH-50\% & Yes & 70\% & \textbf{10.48} & \textbf{14.92} & \textbf{12.70} \\
    & LTH-20\% & Yes & 70\% & 10.99 & 16.17 & 13.58 \\
    \midrule
    \multirow{3}{*}{\textit{\makecell{Dynamic \\ ASR \\ Pathways}}} & LAP-70\% & Yes & 70\% & 10.98 & 16.54 & 13.76 \\
    & LAP-50\% & Yes & 70\% & \textbf{10.82} & \textbf{16.25} & \textbf{13.54} \\
    & LAP-20\% & Yes & 70\% & 10.88 & 16.43 & 13.65 \\
    \bottomrule
  \end{tabular}
  \label{dynamic_pathways}
  }
\end{table}

\subsection{Baselines}
In Table \ref{baselines}, we present the results of existing methods for pruning a multilingual ASR model. We breakdown these methods into three stages: 1) training a dense multilingual ASR model, 2) pruning the dense multilingual ASR model, and 3) training a sparse multilingual model. For reference, we include results of dense monolingual model. Both the IMP and LTH language-specific pruning methods achieve matching performance to the original dense multilingual model and surpass the dense monolingual models. The ASR Pathways method outperforms other methods using the language-specific masks obtained in Stage (2), promoting parameter sharing among languages.

\subsection{Adaptive masking in monolingual pruning}
In the last two rows of Table \ref{baselines}, we present the results of using adaptive masking for monolingual pruning. Our proposed Stage (2) modified the IMP and the LTH language-specific pruning methods in Stage (2) and achieved a consistent 5.3\% relative WER reduction averaged across languages. Comparing the adapted sub-networks to the fixed ones, we noticed about an 80\% similarity and a 20\% difference, indicating the effective adaptation occurs within a small part of the pruning masks. Our proposed Stage (2) also outperforms the sparse multilingual model obtained in Stage (3), providing an efficient alternative when storing multiple models is not a concern.

\subsection{Adaptive masking in multilingual pruning}
In Table \ref{dynamic_pathways}, we show the results of a bilingual experiment when using adaptive masking for multilingual pruning. We initialized the training with the LTH or the LAP masks at the target sparsity level (70\%) and achieved a consistent improvement when only adaptation is enabled. Notably, adapting the LAP-70\% mask achieves a 12.1\% relative WER reduction, indicating the adaptation step has effectively turned the LAP mask to become more language-specific. We noticed a similar but improved performance when using the LTH-70\% masks, suggesting these masks may be robust  at a high sparsity level. 

\begin{table}[t]
  \caption{WER (\%) results on the MLS test set, utilizing language-specific pruning masks. The proposed approach is compared to an existing method, extending to four languages.}
  \label{t2}
  \centering
  \scalebox{0.8}{
  \begin{tabular}{ c | c | c | c c c c | c }
    \toprule
     Model & Initialization & Sparsity & EN & FR & IT & NL & Avg. \\
    \midrule
    \midrule
    \makecell{ASR \\ Pathways} & LTH-70\% & 70\% & \textbf{13.56} & 10.53 & 17.10 & 16.37 & 14.39 \\
    \midrule
    \textit{\makecell{Dynamic \\ ASR \\ Pathways}} & LTH-50\% & 70\% & 14.84 & \textbf{10.35} & \textbf{16.10} & \textbf{15.15} & \textbf{14.11} \\
    \bottomrule
  \end{tabular}
  }
  \label{4languages}
\end{table}
\footnotetext[1]{The union ratio indicates the ratio between surviving parameters in the union of all masks and the total parameters of the network \cite{yang2023learning}}
\footnotetext[2]{Due to time limitation, this result is inferred at an early checkpoint, subject to a better future improvement}
We observed the best overall performance using mask initialization at a middle sparsity level (50\%) when both pruning and adaptation steps are enabled. For the LTH-50\% mask initialization, our \textit{Dynamic ASR Pathways} method outperformed the respective ASR Pathways baseline with a 5.8\% relative WER reduction. From an analysis, we find this model results in a even lower union ratio\footnotemark[1] (0.34) compared to its baseline (0.36), indicating a better multilingual performance is achieved using even less total effective model parameters. We believe this effect can be attributed to the pruning step introduced in our approach that increases parameter sharing (Section 3.3.2).  For different LAP mask initialization, we noticed consistently a significant performance gain compared to its respective baseline. Further, it is almost matching performance to the ASR Pathways baseline using the LTH-70\% masks, showing a benefit of efficiency with the language-specific pruning rounds eliminated.

In Table \ref{4languages}, we present the extended results of applying \textit{Dynamic ASR Pathways} to pruning for more languages, initializing from the LTH-50\% masks. Our proposed approach outperforms the ASR Pathways baseline with a 2\% relative WER reduction\footnotemark[1] on average across four languages. When considering the performance across FR, IT, and NL, it achieves a notable 5.5\% relative WER reduction. When initializing at a 50\% sparsity level, we saved additional rounds of training and pruning for achieving a target sparsity level, showing the efficacy of applying our approach towards efficient pruning of a multilingual ASR model.

\section{Conclusions}
In conclusion, we proposed an adaptive masking approach for both monolingual and multilingual pruning. In the former case, our proposed method achieved a consistent 5.3\% relative WER reduction averaged across languages and outperformed the sparse multilingual model obtained from going through an additional stage, offering a convenient trade-off between storage and efficiency. In the latter case, we showed the efficacy of our approach in pruning and adapting from different pruning mask initalizations. When initialized from language-agnostic pruning masks, our Dynamic ASR Pathways method showed a consistent and comparable performance to the best performance of the ASR Pathways method that uses language-specific pruning masks, indicating a benefit of efficiency with our approach. When initialized from language-specific pruning masks at a 50\% sparsity level, our Dynamic ASR Pathways method outperforms the ASR Pathways method, ranging from a 2\% to 5.8\% relative WER reduction.  For future work, we want to scale our research of multilingual pruning for more languages and explore the option to make pruning masks learnable.




\bibliographystyle{IEEEbib}
\bibliography{refs}

\begin{thebibliography}{10}

\bibitem{shangguan2021dissecting}
Yuan Shangguan, Rohit Prabhavalkar, Hang Su, Jay Mahadeokar, Yangyang Shi,
  Jiatong Zhou, Chunyang Wu, Duc Le, Ozlem Kalinli, Christian Fuegen, and
  Michael~L. Seltzer,
\newblock ``{Dissecting User-Perceived Latency of On-Device E2E Speech
  Recognition},''
\newblock in {\em Interspeech 2021}.

\bibitem{he2019streaming}
Yanzhang He, Tara~N Sainath, Rohit Prabhavalkar, Ian McGraw, Raziel Alvarez,
  Ding Zhao, David Rybach, Anjuli Kannan, Yonghui Wu, Ruoming Pang, et~al.,
\newblock ``Streaming end-to-end speech recognition for mobile devices,''
\newblock in {\em ICASSP 2019}.

\bibitem{gao2021extremely}
Zhifu Gao, Yiwu Yao, Shiliang Zhang, Jun Yang, Ming Lei, and Ian McLoughlin,
\newblock ``{Extremely Low Footprint End-to-End ASR System for Smart Device},''
\newblock in {\em Interspeech 2021}.

\bibitem{massive_mutlilingual}
Vineel Pratap, Anuroop Sriram, Paden Tomasello, Awni Hannun, Vitaliy
  Liptchinsky, Gabriel Synnaeve, and Ronan Collobert,
\newblock ``{Massively Multilingual ASR: 50 Languages, 1 Model, 1 Billion
  Parameters},''
\newblock in {\em Interspeech 2020}.

\bibitem{tjandra2023massively}
Andros Tjandra, Nayan Singhal, David Zhang, Ozlem Kalinli, Abdelrahman Mohamed,
  Duc Le, and Michael~L Seltzer,
\newblock ``Massively multilingual asr on 70 languages: Tokenization,
  architecture, and generalization capabilities,''
\newblock in {\em ICASSP 2023}.

\bibitem{imp}
Song Han, Jeff Pool, John Tran, and William Dally,
\newblock ``Learning both weights and connections for efficient neural
  network,''
\newblock in {\em NeuraIPS 2015}.

\bibitem{progressive}
Michael~H. Zhu and Suyog Gupta,
\newblock ``To prune, or not to prune: Exploring the efficacy of pruning for
  model compression,''
\newblock in {\em ICLR 2018}.

\bibitem{frankle2018lottery}
Jonathan Frankle and Michael Carbin,
\newblock ``The lottery ticket hypothesis: Finding sparse, trainable neural
  networks,''
\newblock in {\em ICLR 2019}.

\bibitem{shangguan2019optimizing}
Yuan Shangguan, Jian Li, Qiao Liang, Raziel Alvarez, and Ian McGraw,
\newblock ``Optimizing speech recognition for the edge,''
\newblock {\em arXiv preprint arXiv:1909.12408}, 2019.

\bibitem{narang2017exploring}
Sharan Narang, Greg Diamos, Shubho Sengupta, and Erich Elsen,
\newblock ``Exploring sparsity in recurrent neural networks,''
\newblock in {\em ICLR 2017}.

\bibitem{renda2020comparing}
Alex Renda, Jonathan Frankle, and Michael Carbin,
\newblock ``Comparing rewinding and fine-tuning in neural network pruning,''
\newblock in {\em ICLR 2020}.

\bibitem{yang2023learning}
Mu~Yang, Andros Tjandra, Chunxi Liu, David Zhang, Duc Le, and Ozlem Kalinli,
\newblock ``Learning asr pathways: A sparse multilingual asr model,''
\newblock in {\em ICASSP 2023}.

\bibitem{datta2020language}
Arindrima Datta, Bhuvana Ramabhadran, Jesse Emond, Anjuli Kannan, and Brian
  Roark,
\newblock ``Language-agnostic multilingual modeling,''
\newblock in {\em ICASSP 2020}.

\bibitem{ogueji-etal-2022-intriguing}
Kelechi Ogueji, Orevaoghene Ahia, Gbemileke Onilude, Sebastian Gehrmann, Sara
  Hooker, and Julia Kreutzer,
\newblock ``Intriguing properties of compression on multilingual models,''
\newblock in {\em EMNLP 2022}.

\bibitem{kannan19_interspeech}
Anjuli Kannan, Arindrima Datta, Tara~N. Sainath, Eugene Weinstein, Bhuvana
  Ramabhadran, Yonghui Wu, Ankur Bapna, Zhifeng Chen, and Seungji Lee,
\newblock ``{Large-Scale Multilingual Speech Recognition with a Streaming
  End-to-End Model},''
\newblock in {\em Interspeech 2019}.

\bibitem{winata21_interspeech}
Genta~Indra Winata, Guangsen Wang, Caiming Xiong, and Steven Hoi,
\newblock ``{Adapt-and-Adjust: Overcoming the Long-Tail Problem of Multilingual
  Speech Recognition},''
\newblock in {\em Interspeech 2021}.

\bibitem{yu2020gradient}
Tianhe Yu, Saurabh Kumar, Abhishek Gupta, Sergey Levine, Karol Hausman, and
  Chelsea Finn,
\newblock ``Gradient surgery for multi-task learning,''
\newblock {\em NeurIPS 2020}.

\bibitem{shaham-etal-2023-causes}
Uri Shaham, Maha Elbayad, Vedanuj Goswami, Omer Levy, and Shruti Bhosale,
\newblock ``Causes and cures for interference in multilingual translation,''
\newblock in {\em ACL 2023}.

\bibitem{Sun_Shao_Li_Liu_Yan_Qiu_Huang_2020}
Tianxiang Sun, Yunfan Shao, Xiaonan Li, Pengfei Liu, Hang Yan, Xipeng Qiu, and
  Xuanjing Huang,
\newblock ``Learning sparse sharing architectures for multiple tasks,''
\newblock {\em AAAI 2020}.

\bibitem{language-adaptive}
Yizhou Lu, Mingkun Huang, Xinghua Qu, Pengfei Wei, and Zejun Ma,
\newblock ``Language adaptive cross-lingual speech representation learning with
  sparse sharing sub-networks,''
\newblock in {\em ICASSP 2022}.

\bibitem{lin-etal-2021-learning}
Zehui Lin, Liwei Wu, Mingxuan Wang, and Lei Li,
\newblock ``Learning language specific sub-network for multilingual machine
  translation,''
\newblock in {\em Proceedings of the 59th Annual Meeting of the Association for
  Computational Linguistics and the 11th International Joint Conference on
  Natural Language Processing (Volume 1: Long Papers)}, 2021.

\bibitem{foroutan-etal-2022-discovering}
Negar Foroutan, Mohammadreza Banaei, R{\'e}mi Lebret, Antoine Bosselut, and
  Karl Aberer,
\newblock ``Discovering language-neutral sub-networks in multilingual language
  models,''
\newblock in {\em EMNLP 2022}.

\bibitem{fu2022losses}
Yonggan Fu, Yang Zhang, Kaizhi Qian, Zhifan Ye, Zhongzhi Yu, Cheng-I~Jeff Lai,
  and Celine Lin,
\newblock ``Losses can be blessings: Routing self-supervised speech
  representations towards efficient multilingual and multitask speech
  processing,''
\newblock {\em NeurIPS 2022}.

\bibitem{lai2021parp}
Cheng-I~Jeff Lai, Yang Zhang, Alexander~H Liu, Shiyu Chang, Yi-Lun Liao,
  Yung-Sung Chuang, Kaizhi Qian, Sameer Khurana, David Cox, and Jim Glass,
\newblock ``Parp: Prune, adjust and re-prune for self-supervised speech
  recognition,''
\newblock {\em NeurIPS 2021}.

\bibitem{yang2022omni}
Haichuan Yang, Yuan Shangguan, Dilin Wang, Meng Li, Pierce Chuang, Xiaohui
  Zhang, Ganesh Venkatesh, Ozlem Kalinli, and Vikas Chandra,
\newblock ``Omni-sparsity dnn: Fast sparsity optimization for on-device
  streaming e2e asr via supernet,''
\newblock in {\em ICASSP 2022}.

\bibitem{mls}
Vineel Pratap, Qiantong Xu, Anuroop Sriram, Gabriel Synnaeve, and Ronan
  Collobert,
\newblock ``Mls: A large-scale multilingual dataset for speech research,''
\newblock {\em arXiv preprint arXiv:2012.03411}, 2020.

\bibitem{shi2021emformer}
Yangyang Shi, Yongqiang Wang, Chunyang Wu, Ching-Feng Yeh, Julian Chan, Frank
  Zhang, Duc Le, and Mike Seltzer,
\newblock ``Emformer: Efficient memory transformer based acoustic model for low
  latency streaming speech recognition,''
\newblock in {\em ICASSP 2021}.

\bibitem{gulati2020conformer}
Anmol Gulati, James Qin, Chung-Cheng Chiu, Niki Parmar, Yu~Zhang, Jiahui Yu,
  Wei Han, Shibo Wang, Zhengdong Zhang, Yonghui Wu, and Ruoming Pang,
\newblock ``{Conformer: Convolution-augmented Transformer for Speech
  Recognition},''
\newblock in {\em Interspeech 2020}.

\bibitem{park2019specaugment}
Daniel~S. Park, William Chan, Yu~Zhang, Chung-Cheng Chiu, Barret Zoph, Ekin~D.
  Cubuk, and Quoc~V. Le,
\newblock ``{SpecAugment: A Simple Data Augmentation Method for Automatic
  Speech Recognition},''
\newblock in {\em Interspeech 2019}.

\bibitem{liu2023learning}
Chunxi Liu, Yuan Shangguan, Haichuan Yang, Yangyang Shi, Raghuraman
  Krishnamoorthi, and Ozlem Kalinli,
\newblock ``Learning a dual-mode speech recognition model via self-pruning,''
\newblock in {\em 2022 IEEE Spoken Language Technology Workshop (SLT)}. IEEE,
  2023, pp. 273--279.

\end{thebibliography}

\end{document}